\documentclass[figures,cite]{epl}
\usepackage[T1]{fontenc}
\usepackage[latin1]{inputenc}
\usepackage{graphicx}
\usepackage{amssymb}
\usepackage{psfrag}
\usepackage{ae,aecompl}

\title{Quantum diffusion in the quasiperiodic kicked rotor}
\author{Hans Lignier\inst{1} \and Jean Claude Garreau\inst{1} \and Pascal Szriftgiser\inst{1}
\and Dominique Delande \inst{2}}
\shortauthor{H. Lignier {\it et al.}}
\institute{
  \inst{1} Laboratoire de Physique des Lasers, Atomes et Mol\'{e}cules, UMR
CNRS 8523, Universit\'{e} des Sciences et Technologies de Lille,
F-59655 Villeneuve d'Ascq Cedex, France\\
  \inst{2} Laboratoire Kastler Brossel, Tour 12, Etage 1, 4 Place Jussieu, F-75005
Paris, France
}

\begin{document}

\maketitle

\begin{abstract}
We study the mechanisms responsible
for quantum diffusion in the quasiperiodic kicked rotor. We report experimental
measurements of the diffusion constant on the atomic version of the
system and develop a theoretical approach (based on the Floquet theorem)
explaining the observations, especially the
``sub-Fourier'' character of the resonances observed in the vicinity
of exact periodicity, i.e. the ability
of the system to distinguish two neighboring driving frequencies in
a time shorter than the inverse of the difference of the two frequencies. 
\end{abstract}

Quantum chaos is the study of \emph{quantum} systems whose classical
limit is chaotic. A major challenge of quantum
chaos is to understand the mechanisms that make quantum chaos
different from classical chaos. An important difference between classical and quantum
systems is the existence in the latter of \textit{interferences} between
various paths. At long times, a large number of 
complicated trajectories interfere.
One could expect the contributions of the
various paths to have uncorrelated phases, so that the interference terms vanish
in the average
after some time, implying that quantum and classical transport should
be identical. This simple expectation is however too naive, because 
phases of the various paths 
are actually correlated; this is for example the case for the 
kicked rotor. 

The quantum kicked rotor has been extensively studied experimentally
in recent years ~\cite{Raizen_LDynFirst_PRL94,Christ_LDynNoise_PRL98,AP_Bicolor_PRL00,Darcy_QRes_PRL01}.
In its atomic version it consists of a cloud of laser-cooled atoms
exposed to short pulses of a far detuned, standing laser wave, 
corresponding to the Hamiltonian (for the external motion of the atoms)
\begin{equation}
H_{0}=\frac{p^{2}}{2}+\frac{K}{T}\cos\theta\sum_{n}\delta(t-nT).
\label{eq:Kicked}
\end{equation}
where $\theta$ is the $2\pi-$periodic position of the rotor, $p$
the conjugate momentum, $T$ the period and $K$ is proportional to
the strength of the kicks. In the classical limit, this system is
chaotic for $K \gtrsim 1$ ~\cite{Chirikov_ChaosClassKR_PhysRep79},
and the motion is an ergodic diffusion in momentum space
for $K>5$. Dynamical localization
(DL)
is the \emph{suppression} of such a diffusion in the quantum
system by subtle quantum interference effects \cite{casati2}.

DL is expected to disappear if the time periodicity is broken~\cite{Casati_IncommFreqsQKR_PRL89},
as experimentally observed in~\cite{AP_Bicolor_PRL00}.
This can be done simply by adding a second series of kicks with a different
period $rT$, giving 
\begin{equation}
H(r,\lambda)=\frac{p^{2}}{2}+\frac{K}{T}\cos\theta\sum_{n}\left[\delta(t-nT)+\delta(t-nrT-\lambda T)\right]
\label{eq:qp}
\end{equation}
where $\lambda$ is the initial phase between the two kick sequences.
If $r$ is rational, the system is strictly time-periodic and DL takes
place, but it is rapidly destroyed around any rational number.
One way of characterizing this sensitivity to the time-periodicity
is to measure the average kinetic energy $\left\langle p^{2}\right\rangle $
of the system as a function of time (or number of kicks), which allows
to deduce its diffusion constant. In this paper, we present new experimental results
and a theoretical interpretation of the physical mechanism responsible
for the destruction of DL in the vicinity of rational $r$.

Our experimental setup is described in~\cite{AP_ChaosQTransp_CNSNS_2003}.
Cold cesium atoms are produced in a magneto-optical
trap, the trap is turned off, and a series
of short pulses of a far-detuned (13.5 GHz $\sim$ 2600 $\Gamma)$
standing wave (around 65 mW in each direction) is applied. At the end of the pulse series,
pulses of counter-propagating phase-coherent beams perform velocity-selective
Raman stimulated transitions between the hyperfine ground state sublevels
$F_{g}=4$ and $F_{g}=3$. A resonant probe beam is then used to measure
the fraction of transfered atoms, which corresponds to the population in
a velocity class. Repeated measurements allow to reconstruct the atomic
momentum distribution $P(p)$. In the periodic case, one observes, for $t>t_{\ell}$, where
$t_{\ell}$ is the so-called \emph{localization time}, two manifestations
of DL: \emph{i}) $P(p)$ is frozen in a characteristic exponential
shape $P(p)\sim\exp(-|p|/\ell)$, where $\ell$ is the \emph{localization
length}, and \emph{ii}) the average kinetic energy of the system tends
to a constant value, or, equivalently, the diffusion constant 
$D=\lim_{t\rightarrow\infty}\left\langle p^{2}\right\rangle/t$
vanishes.

\begin{figure}
\psfrag{p2}{$\langle p^2/\hbar^2k_{\mathrm{L}}^2\rangle $}
\begin{center}
\twoimages[width=4.5cm,angle=270]{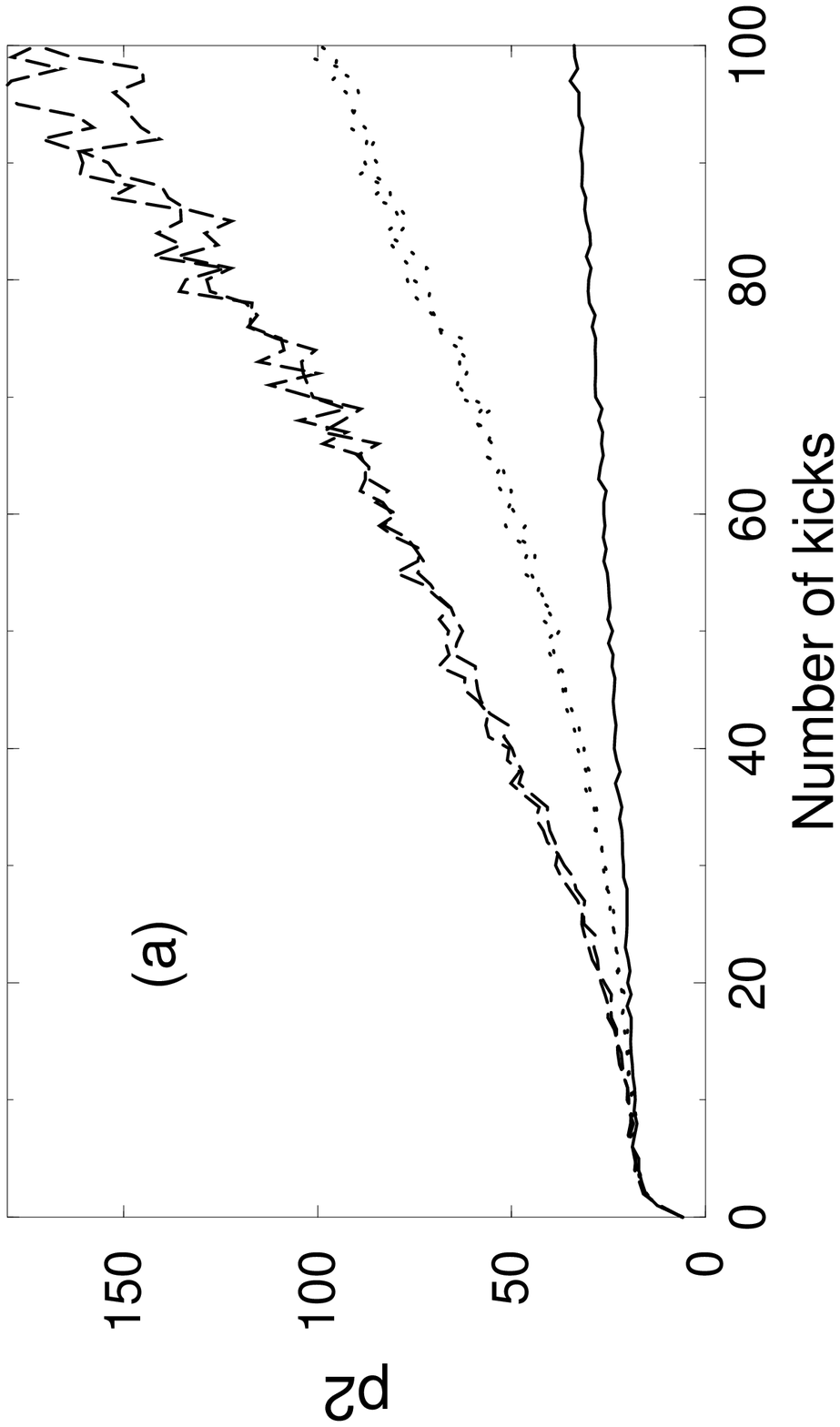}{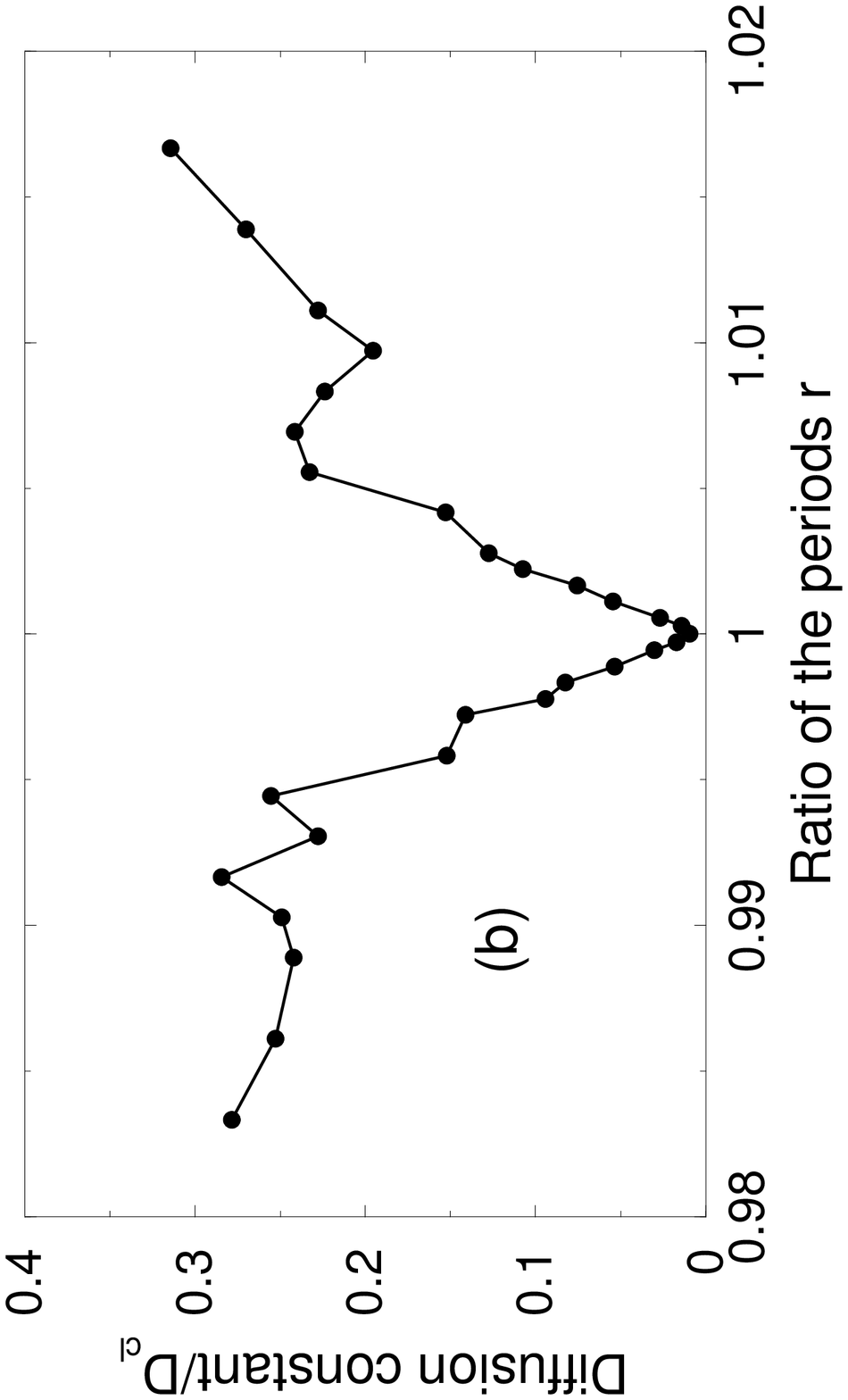}
\end{center}
\caption{(a) Averaged squared momentum $\langle p^{2}\rangle$ ($p$ is in
recoil momentum units) vs. number of double kicks. All curves show
DL after 5-10 kicks. After this break-time, the dynamics is frozen
for the periodic system ($r$=1, solid curve), but a residual quantum
diffusion is observed in the quasi-periodic case, for $r=1+0.00111$
and $r=1-0.00111$ (dotted curves) and $r=1\pm0.00222$ (dashed curves).
(b) The quantum diffusion constant (in units of the classical diffusion
constant) versus $r,$ from data shown in (a). It (almost) vanishes
at $r=1$ because of DL, and rapidly increases symmetrically on both
sides, displaying a triangular cusp.}
\label{fig:diff}
\end{figure}

Experimental measurements of $\left\langle p^{2}\right\rangle $ and
the corresponding diffusion constant are displayed in Fig.~\ref{fig:diff}.
In the periodic case $r=1$ (solid line), after an initial linear
increase, the kinetic energy saturates to a constant value -- corresponding
to $D(r=1)\approx0$ in Fig.~\ref{fig:diff}(b). Due to residual spontaneous
emission by the kicked atoms ~\cite{Raizen_LDynNoise_PRL98}, $D(r=1)$
is not strictly zero. In the quasiperiodic cases, 
there is a residual diffusion due to the quasiperiodicity,
and Fig.~\ref{fig:diff}(b) shows that $D(r)\propto|r-1|$.

\begin{figure}
\begin{center}\includegraphics[width=4.5cm,angle=-90]{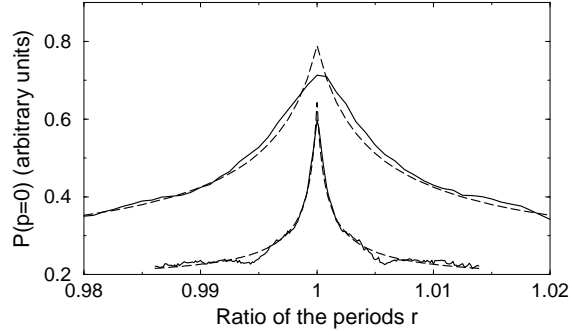}
\end{center}
\caption{Sub-Fourier resonances. The measured atomic population in the zero-momentum
class, after 20 (upper solid curve) and 100 (lower solid curve) double
kicks is plotted vs. the ratio $r$ of the periods. The period of
the first sequence is $T=27.8\mu$s, and $K\approx10.$ At $r=1$,
dynamical localization is responsible for the large number of zero-momentum
atoms. Away from the exact resonance, localization is progressively
destroyed. Note the narrowness (5 times smaller than Fourier limit)
and the triangular shape of the resonance line. Dashed lines are fits
to the experimental lines using Eq.~(\ref{lineshape}).}
\label{pv-r}
\end{figure}

Another way to characterize the observed residual diffusion is to
measure the population $P_{0}$ of the zero-momentum class 
\footnote{As the total number of atoms is constant, measuring $P(p=0)$ is equivalent
to measure the width of the momentum distribution: 
$P(p=0)\propto\left\langle p^{2}\right\rangle ^{-1/2}$.}, 
as a function of $r$. Fig.~\ref{pv-r} shows the experimentally
measured $P_{0}$ after $20$ and $100$ double kicks. This ``localization
resonance'' displays a sharp peak at $r=1$, indicating the presence
of DL, and decreases rapidly on both sides, evidencing the
residual diffusion. The plot presents two surprising features:
(\textit{i}) The resonance is very narrow: after $N$ kicks, it could
be argued that the two quasi-periods can be distinguished only if
they differ by $1/N$ (in relative value). This would predict a width
of the order of $\Delta r=1/N=0.01$ for 100 kicks, whereas we experimentally
observe a width five times smaller, $0.0018$
\footnote{Sub-Fourier factors of 1/37 have been reported in~\cite{AP_SubFourier_PRL02},
and factors of 1/60 have been experimentally observed by us.
}. (\textit{ii}) The ``sub-Fourier'' resonance is not smooth, but
has a marked cusp at the maximum. The understanding of the underlying
mechanism behind the quantum behavior of the system shall also allow
us to explain these features.

Our analysis of the residual diffusion takes the periodic case as the
reference system, because, being periodic, it can be analyzed using
the \emph{Floquet theorem}. A Floquet state (FS) $|\varphi_{k}\rangle$
is defined as an \emph{eigenstate of the unitary evolution operator}
$U(T)$ of $H_{0}$ over one period $T$: 
$U(T)|\varphi_{k}\rangle=\exp\left(-i\epsilon_{k}\right)|\varphi_{k}\rangle$
where $\epsilon_{k}$ is called the eigenphase. 
The temporal evolution
of any state $|\psi\rangle$ after $n$ periods is
$|\psi(nT)\rangle=\sum_{k}{c_{k}\,\,{\mathrm{e}}^{-in\epsilon_{k}}
\,\,|\varphi_{k}\rangle}$
with $c_{k}=\langle\varphi_{k}|\psi(0)\rangle.$ The evolution of
any quantity can be calculated using the basis of FS, for example:
\begin{equation}
\langle p^{2}(nT)\rangle=\sum_{k,k^{\prime}}{c_{k}c_{k^{\prime}}^{\ast}
\,\,{\mathrm{e}}^{-in(\epsilon_{k}-\epsilon_{k^{\prime}})}
\,\,\langle\varphi_{k^{\prime}}|p^{2}|\varphi_{k}\rangle.}
\label{p2}
\end{equation}
The FS of the chaotic kicked rotor are
well-known: they are \emph{on the average exponentially localized
in momentum space} around a most probable momentum $p_{k},$ with
a characteristic localization length $\ell$~\cite{Casati_LocFloquetQKR_PRL90,casati}.
Such a localization -- from which DL originates -- is far from obvious
and is closely related to the Anderson localization in time-independent
disordered one-dimensional systems~\cite{Fishman_LocDynAnderson_PRA84}. 

The initial state is supposed to be 
localized in momentum space around zero-momentum,
with a width much smaller than the width $\ell$ of the FS (which is the case
in the experiment). Hence,
only FS with roughly $|p_{k}|\lesssim\ell$
will play a significant role in the dynamics; we shall call such states
\emph{initially populated} FS. Eq.~(\ref{p2}) is a \emph{coherent}
sum over FS. However, as times goes on, non-diagonal interference
terms accumulate larger and larger phases. In a typical chaotic system,
these phases will be uncorrelated at long enough times, leaving an
\emph{incoherent} sum over FS:
\begin{equation}
\langle p^{2}\rangle\approx\sum_{k}{|c_{k}|^{2}\langle\varphi_{k}|p^{2}|\varphi_{k}\rangle}
\label{v2-DL}
\end{equation}
This equation is valid once DL is established (i.e., for $t>t_{\ell}$).
How long does it take for the phases $n(\epsilon_{k}-\epsilon_{k^{\prime}})$
to be scrambled? This can be simply estimated from the
level spacing between initially populated FS, and turns out to
be roughly $t_{\ell}=\ell T,$ while $\langle p^{2}\rangle$ saturates
to a value $\propto\ell^{2}.$

\begin{figure}
\begin{center}
\includegraphics[width=5.5cm,angle=270]{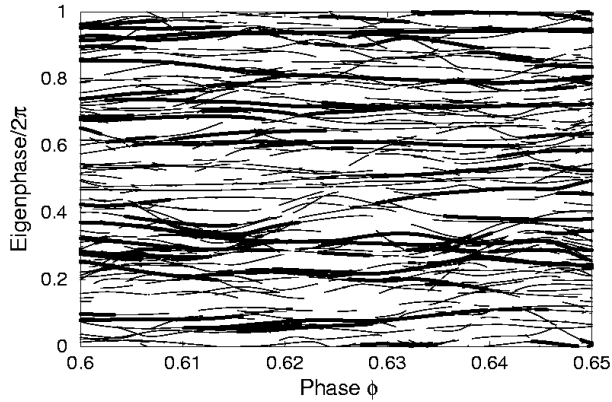}
\end{center}
\caption{Eigenphases $\epsilon_{k}$ of the evolution operator over one period
for the doubly-kicked rotor, corresponding to the experimental conditions,
versus the phase $\lambda.$ The finite duration of the kicks (800
ns) is taken into account. We have only plotted initially populated Floquet states having
a significant weight: $|\langle\psi(0)|\varphi_{k}(\lambda_{0})\rangle|^{2}>$
$10^{-4}$ (thin line), or $>10^{-2}$ (thick line). States appear
(disappear) as their weights go above (below) the threshold. Note
that Floquet states rapidly change when avoided crossings with other
states are encountered.}
\label{level-dynamics}
\end{figure}

A key point is to realize that, if $r$ is close enough to 1, the sequence of kicks is very
similar to a {\em periodic} doubly-kicked rotor for $r=1$, except that
the phase $\lambda$ between the two sequences \emph{slowly} drifts along
the sequence. A small part of the kick sequence around the $n^\mathrm{th}$
kick will seem ``instantaneously'' periodic, with a phase difference
$\lambda(n)=\lambda_0+n(r-1)$, where $\lambda_{0}$ is 
the initial phase between the two kick sequences.
The evolution operator of the quasiperiodic system
from time $(n-1)T$ to $nT$ is thus given by the evolution operator
%$U(\lambda)$ 
$U(\lambda)=\exp(-ip^2\lambda T/\hbar)\exp(-iK\cos\theta/\hbar T)
\exp(-ip^2(1-\lambda)T/\hbar)
 \exp(-iK\cos\theta/\hbar T)$
of the periodic doubly-kicked rotor. The total evolution operator can
thus be written as a product of the ``instantaneous'' evolution
operators $\prod_{n=1}^{N}{U(\lambda_{0}+n(r-1))}.$
For small enough $|r-1|,$ the \emph{adiabatic approximation}
~\cite{Landau} applies: if the system is initially in 
a FS of $U(\lambda_{0}),$ it remains in the corresponding FS
of the  \emph{``instantaneous''} evolution operator as $\lambda$ changes. This
is illustrated in Fig.~\ref{level-dynamics}, which shows the Floquet
spectrum of the periodic doubly-kicked rotor, obtained by numerically
diagonalizing $U(\lambda)$. 
When $\lambda$ is varied the eigenergies evolve along complicated ``spaghetti'', 
characteristic of quantum-chaotic 
systems, with a large number of avoided crossings (AC). 

In the experiment, the initial state is a linear combination of FS.
The adiabatic approximation implies that it remains a linear combination of
the ``instantaneous'' FS with the \emph{same weights} (the phases evolve,
but the squared moduli remain constant). As discussed above, the coherences
between FS vanish after $t_{\ell}$, which implies that Eq.~(\ref{v2-DL})
remains valid close to the resonance, provided one uses the ``instantaneous''
Floquet eigenbasis: 
\begin{equation}
\langle p^{2}(\lambda)\rangle\approx\sum_{k} |c_k|^2
\,\,\langle\varphi_{k}(\lambda)|p^{2}|\varphi_{k}(\lambda)\rangle.
\label{v2-new}
\end{equation}
The adiabatic approximation implies that there are no population exchanges
among Floquet eigenstates. This means that the weights $|c_k|^2$ are constant
all along the evolution; they are evaluated at the initial time, 
corresponding to the initial wavefunction $|\psi (0)\rangle$ and to $\lambda=
\lambda_0$: $|c_k|^2=|\langle\psi(0)|\varphi_{k}(\lambda_{0})\rangle|^{2}$, where
$|\varphi_{k}(\lambda_0)\rangle$ are the ``instantaneous'' eigenstates corresponding to the initial
time. The evolution of the average kinetic energy $\langle p^2 \rangle$ is thus entirely due to the
evolution of the ``instantaneous'' Floquet spectrum with the parameter $\lambda$,
which evolves adiabatically from its initial value $\lambda_0$ to the final value $\lambda=\lambda_{0}+N(r-1)$
corresponding to the end of the kick sequence.
Thus, two distinct types of localization properties come into Eq.~(\ref{v2-new}): those
of the {\em initial} Floquet spectrum, present in the constants $|c_k|^2$, and those of the 
``instantaneous'' Floquet spectra, present in the ``instantaneous'' eigenstates $|\varphi_k(\lambda)\rangle$.
The $|c_k|^{2}$ coefficient gives important weights to the FS represented
in the initial distribution (that we shall call ''initially populated'' FS), which,
because the initial momentum distribution is very sharp compared to the localization
length, are mostly localized around zero momentum. This limits the range of the sum
in Eq.~(\ref{v2-new}) to eigenstates $|\varphi_k(\lambda_0)\rangle$ centered at momenta 
$|p_k(\lambda_0)|<\ell$. 
As $\lambda$ moves away from $\lambda_{0}$, the center $p_k(\lambda)$ of $|\varphi_k(\lambda)\rangle$ 
moves away from zero momentum. As these eigenstates keep the {\em same} weight in the
sum, this {\em enlarges} the momentum distribution. This is the mechanism at the origin
of the quantum diffusion responsible for the growth of the average kinetic 
energy in the
quasi-periodic case. One thus expects $\langle p^{2}(\lambda)\rangle$
to have a minimum at $\lambda=\lambda_{0}$ and to rapidly increase as the
kicks are applied, that is, as $\lambda$ evolves. Since $\lambda-\lambda_{0}$ is proportional
to $r-1$, $\langle p^{2}\rangle$ shall also present a sharp minimum at
$r=1$, which implies that the population in the zero-momentum class
shall present a sharp maximum at $r=1,$ as experimentally observed
in Fig.~\ref{pv-r} 
%\footnote{The sum on the left side of equation
%(\protect \ref{v2-new}) can be numerically evaluated directly. However, 
%the aim of Eq. (\protect \ref{v2-new}) is essentially of providing
%a new insight on the {\em mechanisms} of quantum diffusion, not a calculation
%method for $\langle p^2(t) \rangle$.}

What determines the linewidth of the resonance? A FS may considerably
change because of its AC with other FS, as tiny AC may be crossed
diabatically. On the average, the typical variation $\lambda-\lambda_{0}$
of the phase between kick series for which $|\varphi_{k}(\lambda)\rangle$
loses the localization property of $|\varphi_{k}(\lambda_{0})\rangle,$
is the distance $\Delta \lambda_c$ to the next AC. We immediately deduce that the full
width $\Delta f$ (in frequency) of the sub-Fourier line is such that: 
\begin{equation}
2\Delta\lambda_{c}=(NT)\Delta f.
\label{width}
\end{equation}
 If the classical dynamics is regular, the Floquet eigenstates evolve
smoothly with the parameter $\lambda$; a change in $\lambda$ of
the order of one is thus required to significantly modify the Floquet
states: $2\Delta\lambda_{c}\approx1$, which corresponds to $\Delta f=1/(NT)$,
that is, the Fourier limit. In a classically chaotic system, however,
the level dynamics displays plenty of AC, see Fig.~\ref{level-dynamics},
$\Delta\lambda_{c}$ is then much smaller than unity, leading to 
sub-Fourier resonances. Eq.~(\ref{width}) also predicts the linewidth
to be inversely proportional to the temporal length of the kick sequence
beyond the localization time (i.e. the sub-Fourier character is independent
of $N$), as numerically observed ~\cite{AP_SubFourier_PRL02}. The
critical value $\Delta\lambda_{c}$ depends on the detailed dynamics
of the system. It can be roughly estimated by visual inspection of
the quasi-energy level dynamics, Fig.~\ref{level-dynamics}, to be
of the order of
0.05 and a ``factor 10'' sub-Fourier line, about twice
the experimentally observed factor (the additional
experimental broadening is due to the transverse
profile of the laser mode leading to spatial inhomogeneities in $K$). 

From Eq.~(\ref{v2-new}) and the previous analysis, it is expected
that $\langle p^{2}\rangle$ significantly increases from its minimum
value at $\lambda=\lambda_{0}$. AC between FS localized
around the same momentum are rather large (this is what determines
$\Delta\lambda_{c}$) whereas AC between states localized a distance
$L\gg\ell$ apart in momentum space are typically much smaller and
scale like $\exp(-L/\ell),$ as a consequence of the exponential localization of
the FS. There is thus a very broad distribution
of AC widths, with a large number of tiny AC. A tiny AC typically
extends over a small $\lambda$ interval and thus tends to produce
small values of $\Delta\lambda_{c}.$ The increase of $\langle p^{2}\rangle$
with $\lambda$ thus depends on the number of small AC encountered.
In the presence of exponential localization, the AC density scales
with size $C$ as $1/C$ for $C\rightarrow0$, and 
$\langle p^{2}(\lambda)\rangle-\langle p^{2}(\lambda_{0})\rangle$
shall behave like $|\lambda-\lambda_{0}|$ \footnote{This basically comes from the fact that the probability to
  encounter a ``bad'' AC which expels the atoms from the zero-momentum region is
  proportional to the length of the interval $|\lambda-\lambda_{0}|$.},
producing the cusp experimentally observed in the resonance line,
Fig.~\ref{pv-r}a.
The large number of extremely small AC is responsible for the singularity
of the sub-Fourier resonance line. Another consequence is the diffusive
behavior observed in the vicinity of the resonance, see Fig.~\ref{fig:diff}(b).
Indeed, $\langle p^{2}(\lambda)\rangle-\langle p^{2}(\lambda_{0})\rangle$
increases linearly with $|\lambda-\lambda_{0}|,$ itself a linear
function of time and of $|r-1|.$ Thus, our model correctly predicts
two non-trivial properties: $\langle p^{2}\rangle$ increases linearly
with time and the corresponding diffusion constant is proportional
to $|r-1|.$ This is distinct from the prediction of Random Matrix
Theory, which implies an increase of $\langle p^{2}\rangle$ initially
quadratic in $\lambda-\lambda_{0}$ ~\cite{Altshuler_LevelCorrelation_PRL93}. 

There will always be some degree of nonadiabaticity. Whatever small
$|r-1|$ is, tiny enough AC will be crossed diabatically. This puts
a lower bound on the size of the AC effectively participating in
the quantum dynamics and produces a rounding of the
top of the sub-Fourier resonance line, too small to be seen
in the experiment after 100 kicks, but easily visible after 20
kicks, Fig.~\ref{pv-r}. 

Our approach concentrates on the immediate vicinity of the resonance.
What happens in the wings of the sub-Fourier line? Eq.~(\ref{v2-new})
indicates that this depends on the residual correlation between 
$|\varphi_{k}(\lambda_{0})\rangle$
and $|\varphi_{k}(\lambda)\rangle$ for 
$|\lambda-\lambda_{0}|>\Delta\lambda_c.$
A quantitative answer to this question is difficult. 
However, it seems clear that the quantum diffusion
constant does not exceed the classical one, which corresponds to 
vanishingly small interference terms. Random Matrix Theory tells us that
this type of parametric correlation usually decays algebraically with
$\lambda-\lambda_{0}.$ We thus propose the following \emph{ansatz}:
\begin{equation}
\langle p^{2}(nT)\rangle=\langle p^{2}\rangle_{DL}+D_{cl}\frac{|r-1|}
{|r-1|+\Delta\lambda/t_{\ell}}nT
\label{lineshape}
\end{equation}
 where $D_{cl}$ is the classical diffusion constant and $\langle p^{2}\rangle_{DL}$
the saturation value of $p^{2}$ due to DL. This equation fits (using
$P(p=0)\propto\langle p^{2}\rangle^{-1/2}$) very well the experimental
curves in Fig.~\ref{pv-r}, reproducing the linear behavior at the
center of the resonance and the classical diffusion in the wings.
The parameters are however fitted, not extracted from the previous analysis.
The reason for that is that, in the experiment, the standing wave laser beams creating
the kicking potential have a gaussian profile whose width is only a few
times the width of the spatial atomic distribution. This means that 
atoms away from the center of the beam see a smaller value of $K$ than
the atoms at the center of the beam. There is thus an averaging effect
over $K$ to which the dynamics is sensitive, unfortunately preventing us
form directly comparing the fitted values to the theoretical values of
the parameters.

In summary, we have developed a theoretical approach for the mechanism
of sub-Fourier resonances, which correctly predicts the unexpected
observed features. In particular, our approach evidences the role
of chaotic dynamics for producing narrow sub-Fourier resonances.
Deviations from exact periodicity are treated in
the framework of the adiabatic approximation for the Floquet spectrum
of the system, which thus goes beyond a pertubative approach. The
dynamics is governed by instantaneous Floquet eigenstates, which are
non-trivial objects, as they need to have very well defined internal
(i.e. between the various parts of the wavefunction for each Floquet
state) phase coherence to be stationary states of the periodic
system, but at the same time, the inter-state coherences do not play
any role (beyond the localization time, Floquet states effectively 
add incoherently).
The dynamics is
dominated by an \emph{incoherent} sum of internally extremely
\emph{coherent} states. It shows that the role of interferences in
quantum mechanics is far from obvious, and can produce unexpected
behaviors. 

Laboratoire de Physique des Lasers, Atomes et Molécules (PhLAM) is
Unité Mixte de Recherche UMR 8523 du CNRS et de l'Université des Sciences
et Technologies de Lille. Laboratoire Kastler Brossel is laboratoire
de l'Universit\'{e} Pierre et Marie Curie et de l'Ecole Normale Sup\'{e}rieure,
UMR 8552 du CNRS. CPU time on various computers has been provided
by IDRIS.

\end{document}